\documentclass[conference]{IEEEtran}
\IEEEoverridecommandlockouts
\usepackage{cite}
\usepackage{amsmath,amssymb,amsfonts}
\usepackage{algorithmic}
\usepackage{graphicx}
\usepackage{textcomp}
\usepackage{listings}
\usepackage{xcolor}
\usepackage{hyperref}

\def\BibTeX{{\rm B\kern-.05em{\sc i\kern-.025em b}\kern-.08em
    T\kern-.1667em\lower.7ex\hbox{E}\kern-.125emX}}

\makeatletter
\def\ps@IEEEtitlepagestyle{
    \def\@oddfoot{\footnotesize Presented at Cray User Group 2024 (CUG'24), 2024-05-09.
\hfill}%
    \def\@evenfoot{}%
}
\makeatother

\newcommand{\figref}[1]{Fig.~\ref{#1}}

\begin{document}

\title{GROMACS on AMD GPU-Based HPC Platforms: Using SYCL for Performance and Portability
\thanks{This work was supported by
the Intel Corporation via Intel oneAPI Academic Center of Excellence at KTH,
the Swedish Foundation for Strategic Research,
the Swedish e-Science Research Center,
the National Academic Infrastructure for Supercomputing in Sweden (NAISS 2023/3-27)
and the BioExcel CoE (HORIZON-EUROHPC-JU-2021-COE-01-02),
}
}

\author{\IEEEauthorblockN{Andrey Alekseenko}
\IEEEauthorblockA{\textit{Science for Life Laboratory} \\
\textit{KTH Royal Institute of Technology}\\
Stockholm, Sweden \\
\texttt{andreyal@kth.se} \\
\href{https://orcid.org/0000-0003-4906-7241}{0000-0003-4906-7241}}
\and
\IEEEauthorblockN{Szilárd Páll}
\IEEEauthorblockA{\textit{PDC Center for High Performance Computing} \\
\textit{KTH Royal Institute of Technology}\\
Stockholm, Sweden \\
\texttt{pszilard@kth.se} \\
\href{https://orcid.org/0000-0003-0603-5514}{0000-0003-0603-5514}}
\and
\IEEEauthorblockN{Erik Lindahl}
\IEEEauthorblockA{\textit{Stockholm University} \\
\textit{KTH Royal Institute of Technology}\\
Stockholm, Sweden \\
\texttt{erik@kth.se} \\
\href{https://orcid.org/0000-0002-2734-2794}{0000-0002-2734-2794}}
}

\maketitle

\begin{abstract}
GROMACS is a widely-used molecular dynamics software package with a focus on performance, portability, and maintainability across a broad range of platforms. Thanks to its early algorithmic redesign and flexible heterogeneous parallelization, GROMACS has successfully harnessed GPU accelerators for more than a decade.
With the diversification of accelerator platforms in HPC and no obvious choice for a well-suited multi-vendor programming model, the GROMACS project found itself at a crossroads. The performance and portability requirements, as well as a strong preference for a standards-based programming model, motivated our choice to use SYCL for production on both new HPC GPU platforms: AMD and Intel.
Since the GROMACS 2022 release, the SYCL backend has been the primary means to target AMD GPUs in preparation for exascale HPC architectures like LUMI and Frontier.
SYCL is a cross-platform, royalty-free, C++17-based standard for programming hardware accelerators, from embedded to HPC.
It allows using the same code to target GPUs from all three major vendors with minimal specialization, 
which offers major portability benefits.
While SYCL implementations build on native compilers and runtimes, whether such an approach is performant is not immediately evident.
Biomolecular simulations have challenging performance characteristics: latency sensitivity, the need for strong scaling, and typical iteration times as short as hundreds of microseconds. Hence, obtaining good performance across the range of problem sizes and scaling regimes is particularly challenging.
Here, we share the results of our work on readying GROMACS for AMD GPU platforms using SYCL,
and demonstrate performance on Cray EX235a machines with MI250X accelerators. Our findings illustrate that portability is possible without major performance compromises.
We provide a detailed analysis of node-level kernel and runtime performance with the aim of sharing best practices with the HPC community on using SYCL as a performance-portable GPU framework.
\end{abstract}

\begin{IEEEkeywords}
molecular dynamics, heterogeneous parallelism, GPU, SYCL, GROMACS
\end{IEEEkeywords}

\section{Introduction}

\subsection{GROMACS}

Molecular dynamics (MD) is a widely-used simulation technique, primarily used in chemistry and biophysics but also in other fields.
Traditionally the algorithm was entirely floating-point limited, which caused Molecular dynamics codes to be among the earliest adopters of GPGPUs \cite{NamdGpu:2005,BangForBuck:2015,BangForBuck:2019}.

GROMACS is a free, open-source molecular dynamics engine aiming for performance, portability, and flexibility. It is one of the most widely used HPC codes world-wide, capable of running on a broad range of hardware and software, from laptops to the largest supercomputer.
GROMACS~2024.0 \cite{Gmx:v2024.0} is the latest major release. It is primarily written in C++17 and contains around 470 thousand lines of code\footnote{Calculated using \texttt{cloc 1.90}; not including dependencies and infrastructure}. It is developed fully openly on GitLab, with code review and automated testing ensuring code quality, correctness, and portability.

\begin{figure*}
    \centering
    \includegraphics[width=0.75\linewidth]{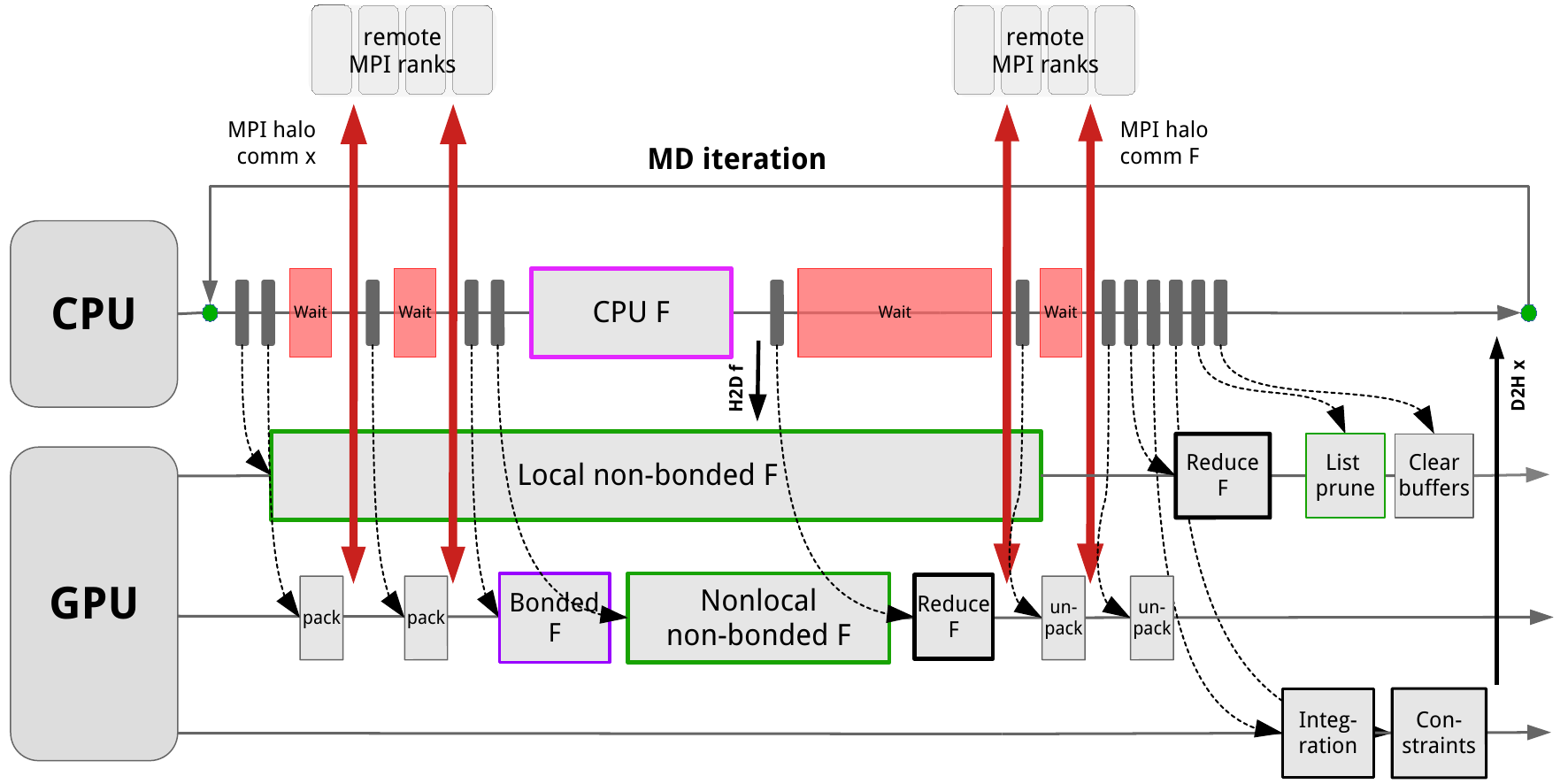}
    \caption{Molecular dynamics task loop with domain-decomposed short-range interaction calculation using direct GPU communication with GPU-aware MPI in a 2D decomposition case with a single pulse in each direction. The grey horizontal arrows show CPU and GPU timelines with boxes as compute tasks or waiting (in red). Grey rounded boxes illustrate GPU kernel launches with dashed arrows pointing to the kernel launched, red vertical arrows represent inter-node communication, solid black vertical arrows show intra-step CPU--GPU data movement only needed with CPU force contributions.}
    \label{fig:dd-md-schedule}
\end{figure*}

Molecular dynamics, in particular biomolecular MD targeting specific large molecules, has the typical performance goal to maximize the number of MD steps for a given simulation system size, making it a strong-scaling problem. Biological MD problems typically span from tens to hundreds of thousands of particles, with a few million particles becoming increasingly common today. This provides a fairly limited amount of data-parallelism when contrasting to the available hardware parallelism in today's accelerated HPC hardware. 
However, since MD timesteps are typically limited to femtoseconds in length, in order to resolve biological timescales, simulations typically involve a very large number of timesteps. 
Improving time-to-solution pushes for ever faster iteration rates, making the problem extremely latency sensitive already at moderate scale. 
On modern hardware, this often translates to sub-millisecond wall-time per MD step executing dozens of compute and intra-/inter-node data movement tasks, including computing forces, halo communication, integration, as shown in \figref{fig:dd-md-schedule}. Even on commodity hardware, modern MD codes run at hundreds of microseconds per iteration at peak, and for high-end hardware iterations can even finish faster than 100 microseconds.\cite{Salomon-Ferrer2013,eastman_openmm_2024,phillips_scalable_2020,GmxPar:2020}

GROMACS uses multi-level parallelism\cite{GmxMain:2015}, targeting each level of hardware parallelism separately. The highest level consists of ensemble-level parallelism, employing multiple simulations with varying degrees of coupling.
Each simulation can be further partitioned across multiple MPI ranks using data- and task decomposition. Task parallelization allows different types of interactions to be computed on different MPI ranks and it is the key ingredient for the heterogeneous parallelization\cite{GmxPar:2020}. Data-parallelism is used in the neutral-territory domain decomposition for calculating short-range interactions\cite{Hess2008} and partitions the simulation system across MPI ranks; and grid decomposition is used for parallelization of long-range force calculation. Within an MPI rank, OpenMP multi-threading is combined with caching and locality optimized algorithms, and with GPU accelerators some or all tasks can be flexibly offloaded. In addition, to exploit SIMD capabilities of modern CPUs, most algorithms are SIMD parallelized using an internal SIMD library and abstraction layer that allows the kernels to be adapted to all currently available SIMD instruction sets.

The support for heterogeneous computation using NVIDIA GPUs was introduced in GROMACS~4.6\cite{GmxMain:2015,GmxExascale:2015}, and
the latest version of the code can either run 
fully offloaded to GPUs or rely on advanced heterogeneous acceleration techniques to maximize
absolute performance by using all available resources 
on each node, which has been particularly powerful
to combine GPU acceleration with efficient scaling.

Early on, significant effort was devoted to ensuring the portability of the heterogeneous parallelization in GROMACS by adding a portable GPU backend based on the open standard-based API OpenCL. The OpenCL backend was added in 2015 with support for NVIDIA, AMD, and Intel GPUs targeting the latter two for production.
Since the vast majority of GROMACS GPU users were running on NVIDIA hardware, the CUDA backend naturally received more development and performance tuning focus over the years. However, the OpenCL backend has been maintained, regularly tested in CI, as well as expanded despite the relatively small number of users.

As AMD and Intel re-entered the HPC GPU accelerator market with announcements of Frontier (at the Oak Ridge National Laboratory) and Aurora (at the Argonne National Laboratory) exascale systems, the GPU portability in GROMACS could no longer be on a ``best-effort'' basis. Yet, despite the proliferation of heterogeneous programming models, it is hard to find the one that is both performance and fully supported by all vendors\cite{GPU-Programming-Model-Overview}. AMD's ROCm software stack uses the HIP language heavily inspired by CUDA. The HIP compiler and runtime are open-source, but their development process is not open and, like CUDA, the language is not based on a standard. While HIP is intended as a portable solution, that portability is limited to NVIDIA GPUs.
Our interest in Intel's oneAPI software stack with Data Parallel C++ (DPC++) is based on SYCL, which \emph{is} an open standard-based programming model, making it attractive as a performance-portability layer, in particular with multiple open-source implementations available.

For GROMACS to target the new exascale heterogeneous platforms a number of key ingredients were required: scalable algorithms, a suitable GPU runtime, and a few external libraries. 
Future-proof GPU-native algorithms, accelerator-resident parallelization, and scalable direct-GPU communication schemes have been a long-term focus, and their development continues.
Since we expect increasing use of accelerator-dense node architectures with fast GPU-GPU interconnects and optimized GPU-NIC data-paths, to minimize CPU-GPU data movement all key algorithms have been ported to GPUs, and we also implemented direct GPU-GPU communication.
At the same time GROMACS maintained its heterogeneous design and the ability to use the available CPU resources concurrently with GPU execution for performance or in support of the broad GROMACS features set\cite{GmxPar:2020}.
 External math and communication libraries are also critical for performance. Fast Fourier transforms (FFTs) are needed by the particle mesh Ewald algorithm (PME) used for computing long-range interactions. PME requires multiple 3D FFTs per time-step, with data distributed across ranks. Transform sizes characteristic to biomolecular simulations are relatively small (tens- to hundreds of points per dimension).
PME scaling is a major challenge on current heterogeneous HPC systems and requires a distributed FFT library optimized to scale even for small transforms. GROMACS' scalability with PME decomposition has been demonstrated using the cuFFTmp library for NVIDIA GPUs\cite{cuFFTMp-blog:2022}, but no comparably scalable option exists for AMD GPUs. GROMACS integrated heFFTe\cite{HeFFTe:2020} as a portable FFT backend, which has AMD GPU support using rocFFT. However, as heFFTe lacks the scalability for small 3D FFTs, scaling beyond a single node is limited, and only possible when simulating a very large system\cite{Kronberg:poster:2024}.
For direct GPU communication GROMACS uses GPU-aware MPI. While MPI lacks GPU queue-awareness (requiring additional synchronization on the critical path) and CPU-initiated communication has latency penalties, this is currently the most robust and portable approach, in particular for AMD and HPE Cray systems.

\subsection{SYCL}

SYCL is a C++17-based standard for heterogeneous computing that is royalty-free and offers a platform to write vendor-agnostic performant code. Unlike alternatives like HIP or Kokkos, SYCL's status as an industry standard means that it comes with future compatibility assurances, and while its early development has certainly depended on efforts contributed by vendors, nobody can limit the community's ability to port, improve, and adapt the language.

\subsubsection{SYCL programming model}

The SYCL design has been inspired by OpenCL but uses a single-source model and allows using modern C++ features in the kernels. SYCL's execution model is similar to those of other direct GPU programming frameworks, like CUDA, OpenCL, or HIP: a kernel (a functor or a lambda function in SYCL) is executed over a multidimensional range. The range is divided into work-groups, which are further divided into work-items, with each work-item executing the same kernel code.

With a few exceptions discussed below, SYCL serves as a thin portability layer on top of the native vendor API. When running on AMD GPUs, this means that an in-order \texttt{sycl::queue} corresponds to a \texttt{hipStream}, and \texttt{sycl::malloc\_device} uses \texttt{hipMalloc} internally. This entails that, for example, profilers (like \texttt{rocprof} and \texttt{nsys}) and debugging tools (like \texttt{rocgdb}) handle SYCL applications just like the native ones.

The SYCL standard supports two distinct approaches to data management and scheduling. The first one, ``buffer-accessor'' approach, relies on defining data dependencies between different operations. It allows automatic data movement and synchronization and aims to simplify heterogeneous programming, but, in practice, SYCL runtimes are unable to match the performance of manually-scheduled code, and high-performance applications rarely use it. The second approach, based on unified shared memory (USM) and in-order queues, requires manual memory allocation and copies, as well as explicit synchronization between different asynchronous operations. This approach is similar to what CUDA and HIP use and typically allows achieving higher performance.

SYCL has been designed with interoperability in mind from the start. It provides a family of \texttt{sycl::get\_native} functions to obtain the underlying native backend objects from SYCL objects. While harming portability, it enables seamless interoperability with native vendor-provided libraries, which is crucial for an ecosystem in its early stage. In GROMACS, this enables the use of native FFT and GPU-aware MPI libraries with only a thin wrapper to convert between SYCL and native objects.

\subsubsection{Implementation and adoption}

For targeting high-performance GPUs, there are currently two relevant SYCL implementations: oneAPI DPC++, developed by Intel and Codeplay, and AdaptiveCpp (previously known as hipSYCL and Open SYCL), developed by a research group at Heidelberg University.
There are other SYCL implementations, but they are either less mature or target different hardware.

Intel oneAPI DPC++\cite{DpcppBook:2023} is a fork of the LLVM Clang compiler\footnote{\url{https://github.com/intel/llvm/}} extended with SYCL compilation passes. Binaries can be downloaded from the Intel website or built from source. Initially, oneAPI DPC++ only supported CPUs and Intel GPUs, but later, in collaboration with Codeplay, support for NVIDIA and AMD GPUs was added based on the CUDA driver API and HIP API, respectively.

AdaptiveCpp\cite{AdaptiveCpp:2022,AdaptiveCpp:2020} is implemented as a Clang plugin and wrapper scripts. It started as a layer on top of HIP (hence its original name, hipSYCL), but later added support for NVIDIA GPUs via CUDA runtime API and arbitrary OpenCL GPUs. Notably, AdaptiveCpp is able to create application binaries compatible with future AMD GPU architectures by embedding kernels as a portable bytecode\cite{AdaptiveCpp:2023}, something not possible even with the AMD ROCm toolchain.

Both these implementations leverage the mature and well-developed open-source LLVM compiler infrastructure, which is also the basis for most vendor compilers today, including the AMD HIP compiler. This means that both oneAPI DPC++ and AdaptiveCpp have the same hardware support guarantees as the native API. 

The runtime built upon the native API enables the use of any tool supporting the native vendor toolchain, such as AMD's \texttt{rocprof} or NVIDIA's \texttt{compute-sanitizer}, with SYCL applications. And the interoperability support in the SYCL standard makes it easy to integrate the use of native libraries: both high-level, like FFT or BLAS, and low-level ones, like GPU-aware MPI or profiler annotation API.

\subsection{The evolution of SYCL support in GROMACS}

The work on GROMACS SYCL backend started in 2020. The initial focus was on enabling Intel GPUs, but support for building with AdaptiveCpp (then hipSYCL) was added early as an experiment in portability to different hardware and as a way to ensure standard-compliance and to avoid unintentionally relying on oneAPI-specific behaviors\cite{GmxSycl:2021}.

GROMACS 2022 extended the range of supported kernels and selected AdaptiveCpp (then called hipSYCL) as the recommended backend for targeting AMD GPUs. The primary focus was data center devices like the Instinct MI250X, but consumer devices were also supported to the extent they are supported in ROCm. With the OpenCL backend deprecated, SYCL was chosen over AMD ROCm in order to reduce code duplication and thus maintenance costs while also allowing faster enablement of new features across all the supported hardware. The SYCL backend in GROMACS~2022 supported all main force kernels and integration/constraints to allow GPU-resident mode with both Intel oneAPI DPC++ and AdaptiveCpp on GPUs from all three major vendors\cite{PascPoster:2022}.

In GROMACS 2023, the listed forces kernel was ported. Another major improvement was direct GPU communication using GPU-aware MPI, which was crucial for efficient scaling across multiple GPUs, especially on systems like Cray EX235a with NICs directly attached to GPUs.

Since GROMACS is one of the first complex codes to embrace SYCL on AMD GPUs, it has encountered a unique set of challenges. Similar problems are likely to arise when other applications attempt to strong-scale and require fine-grained parallelism. Hence, the solutions discussed are aimed to serve as guidance for others eager to use SYCL on the current and future generation of supercomputers.

We start by characterizing the kernel and end-to-end application performance on a single GPU for a range of system sizes and discuss how various settings of the SYCL runtime affect its performance. Then we extend the testing to study the strong-scaling performance on multiple nodes. Finally, we compare the performance of GROMACS with the SYCL backend with an independent HIP fork on widely used benchmarks, both when running on a single GPU and when scaling up to 512 nodes.

\section{Materials}

\subsection{HPC systems and programming environments}

The performance was measured on two HPE Cray EX supercomputers. Both use Cray EX235a blades, with 64-core AMD EPYC 7A53 CPU and four AMD Instinct MI250X accelerators in each, and HPE Slingshot interconnect.

\begin{itemize}
    \item LUMI: The LUMI-G partition consists of 2978 Cray EX235a nodes. Cray PE~23.09 version was used, together with ROCm~5.4.6, \texttt{libstdc++}~12.2.0. One CPU core on each core compute complex (CCX) is reserved and SMT is disabled by default, so user jobs have 7 cores (7 hardware threads) available per GCD.
    \item Dardel: The GPU partition consists of 62 Cray EX235a nodes. Cray PE~22.06 version was used, together with ROCm~5.3.3, \texttt{libstdc++}~12.2.0. No CPU cores are reserved and SMT is enabled by default, so user jobs have 8 cores (16 hardware threads) available per GCD.
\end{itemize}

Note that each AMD Instinct MI250X GPU contains two Graphics Compute Dies (GCD), and each GCD has its own HBM memory and is presented by the AMD ROCm stack as an individual GPU. Therefore, each node with four MI250X modules provides, from the application perspective, 8 GPUs.

\subsubsection{Affinity configuration}

Cray EX235a nodes have a complex topology\footnote{\url{https://docs.lumi-supercomputer.eu/hardware/lumig/}} which requires optimizing the hardware assignment to application processes and threads.
The CPU is composed of 8 Compute Core Complexes (CCX), each with 8 cores and a private L3 cache. Each CCX is connected to a central I/O hub which in turn links to the Graphics Compute Dies (GCD) using Infinity Fabric. While important technical details of the I/O die internal topology are not disclosed, each CCX is documented to have affinity to a single GCD. All GCDs in a node are also interconnected with Infinity Fabric (which is non-uniform).
Finally, each GPU is connected to a Cray Slingshot NIC. This is further complicated by the different ordering of different components: e.g.~CCX~0 (cores 0--7) is connected to GCD~4 which is connected to NIC~2. 

The GROMACS OpenMP multi-threaded algorithms are optimized for data accesses to take advantage of adjacent thread placement. The AdaptiveCpp runtime also uses multiple worker threads, and we observed its performance to degrade when threads are placed in different L3 domains. As a latency-sensitive workload, GROMACS wants to minimize the number of hops from CPU to GPU and from GPU to NIC.

In general, our affinity protocol aims to keep all threads of an MPI rank, both application and runtime, bound to a set of cores or hardware threads, here that of a CCX. Affinity of application threads is controlled per-thread to maintain ordering and cache locality. Furthermore, we optimize for CCX-GCD NUMA distance by reordering either GPU IDs (Dardel) or reordering the MPI rank to CCX mapping (LUMI).
To achieve this, on Dardel we used \texttt{--cpus-per-task=16 --cpu-bind=cores} SLURM options to pin each MPI rank to a single CCX and a wrapper script setting per-rank the \texttt{ROCR\_VISIBLE\_DEVICES} environment variable to select the matching GCD. 
On LUMI we used \texttt{--cpu-bind=mask\_cpu:...} custom binding 
to reorder MPI ranks and map them to the seven cores available to the application, combined with a simple sequential \texttt{ROCR\_VISIBLE\_DEVICES=\$SLURM\_LOCALID} rank to GPU mapping. Note that since a job cannot bind any threads, not even HSA or MPI threads to the reserved core, this configuration is unfavorable to multi-threaded applications that use the CPU since just seven cores (with SMT off) need to be shared by application threads with multiple runtime threads.
To ensure the adjacent placement of application threads we used \texttt{OMP\_PLACES=cores} and \texttt{OMP\_PROC\_BIND=close} environment variables. The NIC affinity was controlled by using \texttt{MPICH\_OFI\_NIC\_POLICY=GPU} setting of Cray MPICH.

\subsection{Software versions}

\begin{itemize}
    \item GROMACS~2024.1\cite{Gmx:v2024.1} and 2024.0\cite{Gmx:v2024.0}: The most recent  version of GROMACS using SYCL to target AMD GPUs. There are a number of bugfixes in the 2024.1, including a fix for a CPU performance regression, but none of them are expected to affect the GPU performance.
    \item GROMACS~HIP: Independently from the SYCL backend development, AMD and Stream~HPC made available a fork of GROMACS based on GROMACS~2022.beta2\footnote{\url{https://github.com/ROCm/Gromacs/commits/9f01a09a4356/}}. It used a semi-automated approach to convert the GROMACS CUDA backend to HIP with multiple manual optimizations. This version does not include fixes for correctness issues identified in mainline GROMACS after December 6, 2022.
\end{itemize}

The GROMACS HIP fork was used in earlier benchmarking\cite{EarlyMdAmd:2022} and we use it for performance comparisons in this paper, but it should be kept in mind that its performance and behavior cannot be directly compared to the upstream GROMACS due to significant divergence of the two codebases over the two years since the fork. The HIP fork includes a number of workarounds for AMD GPU compiler issues not present in the upstream GROMACS, as well as more MI250X-specific tuning. The upstream GROMACS, on the other hand, has general performance improvements to MPI communications, as well as multiple bugfixes.

We also compare the performance of two versions of the AdaptiveCpp runtime: 0.9.4\cite{AdaptiveCpp:tag:0.9.4} and 23.10.0\cite{AdaptiveCpp:tag:23.10}.
In addition to the project changing name from hipSYCL to AdaptiveCpp to reflect its broader scope, AdaptiveCpp~23.10.0 received numerous improvements to the SYCL runtime library to address latency-sensitive workloads, reducing its CPU overheads and adding a new scheduling mode.
Both versions were compiled using the \texttt{amdclang++} compiler from ROCm toolchain against the LLVM from the same toolchain. The multipass HIP compiler mode was used when building GROMACS with AdaptiveCpp.

The GROMACS~HIP version was built using Cray compiler wrappers (\texttt{CC}) with the same ROCm/HIP device compiler as used for the SYCL build.

\subsection{Benchmark systems}

We employ the following benchmark systems to test GROMACS performance and scaling.

\begin{itemize}
    \item \textit{Grappa PME}: A set of artificial benchmark systems, containing a mix of ethanol and water molecules, set up to use long-range (PME) interactions. The size of the system ranges from 1~500 atoms to 46 million atoms. Despite limitations (e.g., they have uniform density, which is rarely the case for biomolecular systems studied using MD), they are fairly representative of a typical computational load for atomistic systems. These systems are used to study performance on a single GCD for different system sizes.
    \item \textit{Grappa RF}: same as Grappa PME but using short-range reaction-field electrostatics. While atomistic simulations typically use PME, the reaction-field electrostatics are used in coarse-grained Martini \cite{Martini:2023} forcefield, which is commonly used for simulating extremely large systems \cite{Martini-132M-beads:218}. These systems are used to study strong scaling up to 512~nodes since their scaling is governed solely by domain decomposition and not limited by a single PME rank.
    \item \textit{STMV}: Satellite Tobacco Mosaic Virus is a small plant virus. This is a common benchmark system for molecular dynamics applications, representing a moderately-large atomistic system. It contains 1~066~628 atoms and uses PME electrostatics. It is used to study strong scaling within a single node (up to 8 GCDs).
    \item \textit{ADH}: A protein enzyme solvated in a box of water. The Cubic version uses a cubic box and contains 134~177 atoms, while the Dodec version uses a dodecahedral box and contains 95~561 atoms. Both versions use PME electrostatics. It is a representative average-sized protein system.
    \item \textit{RNAse}: a small nuclease protein solvated in a box of water. The Cubic version uses a cubic box and contains 24~024 atoms, while Dodec version uses a dodecahedral box and contains 16~816 atoms. Both versions use PME electrostatics. It represents a small protein system.
\end{itemize}

\section{Performance characterization}

\subsection{Kernel performance}

\begin{figure}
    \centering
    \includegraphics[width=3.4in]{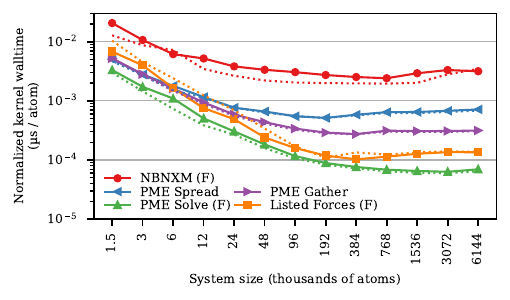}
    \caption{Absolute kernel walltime per atom for Grappa PME, as a function of the system size for selected force kernels. Solid lines show data for GROMACS~2024.0 with AdaptiveCpp~23.10.0; dashed lines show the performance of the HIP fork.}
    \label{fig:kernels-absolute}
\end{figure}

The most compute-intensive part of molecular dynamics is force computation. Other tasks, such as integration, are rarely the limiting factor for performance and are primarily run on the GPU for data locality and to minimize data transfers. Figure~\ref{fig:kernels-absolute} shows the single-GCD force kernel timings (wall-time microseconds per atom) for GROMACS 2024.0 with AdaptiveCpp~23.10 (solid lines) and GROMACS~HIP (dashed lines) when run on a single GCD, as reported by \texttt{rocprof} with kernel and memory operations serialized.

The left side of the graph is the regime where the algorithm does not have sufficient data/compute to saturate the hardware. This means low amounts of data per wave-count, and limited utilization of the GPU relative to the startup/wind-down, which means the execution is effectively limited by latency.  
Even the \emph{longest-running} kernel, NBNXM (short-range non-bonded forces), has a walltime as low as 19.2~\textmu{}s (in the GROMACS~HIP version). As the number of atoms in the system increases, the GPU gets saturated and the performance per atom plateaus after 384 thousand atoms.
But even at the right side of the graph, when running a 6~million-atom system, the NBMXM kernel has a walltime of only 20~ms and other kernels have sub-millisecond walltime.

This also shows that the NBNXM kernel is the rate-limiter for single-GPU performance, with PME Spread and Gather kernels following. The PME Solve kernel is on the critical path for long-range electrostatics (PME) computation, and thus is also of importance. The Listed Forces kernel is relatively inexpensive and can usually overlap well with the local NBNXM kernel, at least when using domain decomposition (see \figref{fig:dd-md-schedule}).

When comparing the kernel performance, it is important to note differences between GROMACS~2024 and GROMACS~HIP, both in the kernel code and in the related algorithms:
\begin{itemize}
    \item GROMACS~2023/2024 include a correction to limit pressure deviations which can affect the
    pair list buffer and therefore the NBNXM pair kernel performance which is missing from the HIP fork due its old base.
    For the systems compared, the difference in the list range is usually minor (1.1\% for Grappa PME, 0.1\% for STMV) and the performance of the affected kernels is not significantly impacted. 
    \item GROMACS~HIP implements a post-pruning list sorting strategy aimed at further reducing tail effect of NBNXM and prune kernels. This can noticeably affect kernel performance for smaller system sizes, at the cost of launching additional kernels to perform the sorting on the GPU. This optimization is not yet implemented in the SYCL backend of GROMACS. Based on the improvement it provides in HIP fork, we estimate 10-25\% walltime reduction for NBNXM kernels with small system sizes. On the flip side, the list sorting increases the complexity of the Prune-only kernel as it now has an additional task of computing the list size histogram.
    \item GROMACS~HIP contains a number of tweaks aimed at working around compiler issues and improving code generation. While most of these have been added to the SYCL backend, some proved to be complex with mixed benefits and were therefore not ported in the hope that future AMD compilers will fix these issues.
    Such device-specific code divergence poses a long-term maintenance burden, in particular since the GROMACS kernels make extensive use of kernel specialization leading to a range of flavors, each potentially impacted differently by such tweaks. 
    \item The GROMACS~HIP LeapFrog kernel, in addition to integration, also clears the PME grid. In upstream GROMACS, this is done by a separate \textit{memset} call.
    \item Different GROMACS versions use slightly different versions of the VkFFT\cite{VkFFT} library for doing FFT on GPU:
    GROMACS~2024 uses VkFFT~1.3.1,
    and GROMACS~HIP uses a forked version with additional optimizations (only partially upstreamed into VkFFT~1.3.1).
\end{itemize}

\begin{figure}
    \centering
    \includegraphics[width=3.4in]{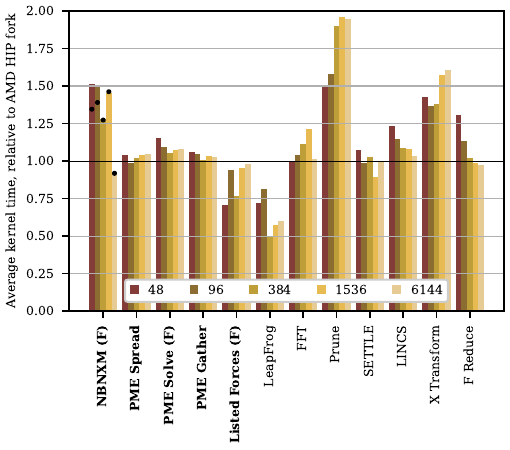}
    \caption{Kernel time for selected kernels in GROMACS~2024.0 built with AdaptiveCpp~23.10.0, relative to the timing of respective kernels in the HIP fork. Differently colored bars correspond to various input sizes (48, 96, 384, 1536, 6144 thousand atoms). Bold labels indicate force kernels presented in \figref{fig:kernels-absolute}.\\
    Note that the HIP fork contains a pair list sorting optimization, not yet implemented in the SYCL version. This affects NBNXM kernel performance for small inputs; black dots show the performance of the SYCL kernels relative to the HIP fork with pair list sorting disabled.\\
    The Grappa PME systems use ``tabulated electrostatics, Lennard Jones with combination rule'' version of the NBNXM kernel. The performance varies for different NBNXM kernel types.
    } 
    \label{fig:kernels-relative}
\end{figure}

In \figref{fig:kernels-relative}, one can see the performance of GROMACS~2024.0 SYCL kernels relative to the kernels in the HIP fork for a range of Grappa PME systems when running on a single GCD.

While the performance of GROMACS~2024.0 SYCL kernels is usually slightly lower than the tuned GROMACS~HIP kernels, in many cases they are very close (e.g., SYCL PME Spread kernel time is between 1\% faster and 7\% slower than HIP kernel, depending on the system size; and Listed Forces SYCL kernel is faster than its HIP counterpart). However, the performance of the NBNXM kernel is lagging behind. This is partially due to the lack of cluster sorting in the SYCL backend; the black dots in \figref{fig:kernels-relative} show the performance of unmodified GROMACS~2024.0 relative to GROMACS~HIP with pairlist sorting disabled. Another factor is that the NBNXM kernels are relatively complex, have high register pressure hence sensitive to register allocation, and are very sensitive to optimal code generation which has proven to be a major challenge for the AMD GPU compiler.

Importantly, this shows that \emph{there are no kernel performance limitations inherent to SYCL} relative to the native API.
GROMACS' architecture is flexible enough to allow instantiating different kernel options for different hardware (e.g., to optimize for different sub-group / warp sizes) and to allow specializing kernels for particular devices using preprocessor conditionals (e.g., to call device-specific intrinsic or to control the loop unrolling to optimize for a particular architecture). Such optimizations are usually not a maintenance burden when they can either be expressed generically (e.g., an integer parameter controlling algorithm unrolling) or are very local (replacing a standard function call with an intrinsic).
Specialization in a portable codebase becomes a challenge when it diverges enough from the ``main'' code that one needs to introduce significant specialized code chunks in several places of the codebase; in such a case, one has to strike a balance between code unity and performance. Ultimately, it is possible to drop-in complete HIP kernels in a SYCL application: while that will not be portable, it will completely eliminate any possible compiler differences --- and it would enable the code to use hardware features only accessible through a native API.

\subsection{Runtime performance}

While the kernels represent the main performance limiter for larger systems, the bottleneck starts to shift to the runtime and scheduling tasks to the GPU when we enter sub-millisecond iteration times, either when running small systems on a single GPU or in the limit of strong scaling across multiple GPUs.

The AMD HIP API is built on top of the low-level AMD HSA (Heterogeneous System Architecture) runtime, which serves as an interface between userspace applications and the OS kernel module. This approach requires the use of a separate worker thread running an event loop to handle HSA signals. In our tests with ROCm~5.3.3 and earlier, this worker thread has CPU utilization around 70--80\%, meaning that at least one CPU (a full core or a hardware thread on systems with SMT) should be reserved for it. Notably, the default behavior of the HSA thread is to reset the affinity mask inherited from the parent process, disregarding the affinity set by CPU pinning tools\footnote{The cores reserved on LUMI-G for \emph{low-noise mode} are not accessible to this thread}. This behavior can be countered by setting an undocumented environment variable \texttt{HSA\_OVERRIDE\_CPU\_AFFINITY\_DEBUG=0} (available since ROCm~5.3.0), making the HSA thread follow the typical affinity setting process. In our observations, this behavior does not significantly alter the performance, but makes scaling measurements more predictable by making CPU contention more controllable.

AdaptiveCpp works as an additional layer on top of the HIP runtime and supports two modes of operation: the default (asynchronous) behavior and the instant submission mode. In both cases all ``management'' API calls, such as memory allocation or stream synchronization, directly call the underlying HIP API. The difference between the two modes is in how kernels and other asynchronous tasks are handled.

In the default mode, when the application uses the \texttt{sycl::submit} API to schedule a kernel, a node is added to the internal AdaptiveCpp task graph, but no HIP API calls are made. When a synchronization point is reached (e.g., \texttt{sycl::event::wait} call analogous to \texttt{hipEventSynchronize}) or when a graph size passes a certain node count threshold (set via \texttt{HIPSYCL\_RT\_MAX\_CACHED\_NODES} environment variable, by default 100), the graph is \emph{flushed} to a special worker thread which issues the necessary HIP API calls. A second worker thread is used to monitor the status of the submitted operations and free or recycle unused runtime objects. The purpose of this scheme is to enable advanced SYCL scheduling mechanisms, such as out-of-order queues and multi-device queues with automatic load balancing. Additionally, the kernel submission call, from the application perspective, can be faster than when calling HIP directly.

The instant submission mode was introduced in AdaptiveCpp~23.10.0 and is aimed at being closer to metal. In this mode, the \texttt{sycl::submit} directly calls the HIP API internally, making the runtime operations much more similar to using the native vendor API. The instant submission mode is, by design, not compatible with any advanced scheduling approaches, and aimed at the applications that have the CUDA-like approach to working with the GPU.

\subsubsection{Recording events}

The SYCL standard states that every enqueued operation must return a \texttt{sycl::event} object which can be used to check operation completion (directly or indirectly by using it to signal the dependency of subsequent operations on the current one). When SYCL is implemented on top of HIP, both oneAPI DPC++ and AdaptiveCpp use \texttt{hipEventRecord} (and, potentially, \texttt{hipEventCreate} and \texttt{hipEventDestroy}) to obtain this synchronization point. This means that submitting a kernel or any other asynchronous operation requires not one, but two (or even four) HIP API calls. This introduces additional latency to the GPU submission path: a single \texttt{hipEventRecord} call takes, on average, around 2~\textmu{}s, but, in ROCm~5.3.3 and earlier, it can occasionally take over 30~\textmu{}s, as reported by \texttt{rocprof}. Additional events also increase the workload of both the underlying HSA runtime (producing extra barrier packets) and the top-level AdaptiveCpp runtime (which manages the pool of \texttt{hipEvent} objects). However, GROMACS mainly relies on in-order queues and discards most of the events returned from SYCL API calls, meaning that these extra overheads serve no purpose. While the SYCL standard does not have any facility for the required optimizations, AdaptiveCpp offers a \textit{coarse-grained events} extension, which allows the application to selectively relax the synchronization guarantees of returned \texttt{sycl::event} objects, allowing the SYCL runtime to avoid the API calls for event recording and management\footnote{The same problem affects oneAPI DPC++, but their extension, ``discard events'', is not currently suitable for GROMACS use case.}.

\begin{figure}
    \centering
    \includegraphics[width=3.4in]{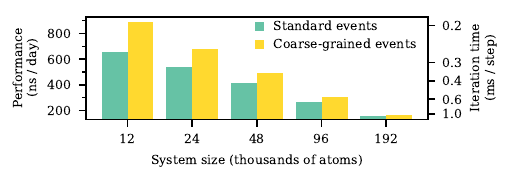}
    \caption{Total application performance (simulated nanoseconds per day, more is better) and corresponding average iteration rate (walltime milliseconds per step) for a range of Grappa PME boxes, run using AdaptiveCpp (hipSYCL) 0.9.4 and GROMACS~2024.0 with and without coarse-grained events optimization.}
    \label{fig:grappa-cg-plot}
\end{figure}

For latency-bound cases (less than 20~000 atoms per GCD), this optimization alone leads to over 35\% speed-up in the total application performance, as can be seen in \figref{fig:grappa-cg-plot}. The effect is diminished for larger systems as the kernel submission latency becomes less of a bottleneck, but even for 192~000 atoms per GCD the difference is noticeable (7.9\%).

The coarse-grained events extension is used in all GROMACS versions since 2023, and its effect is not further evaluated, as there are no known downsides to using it given GROMACS scheduling approach.

\subsubsection{AdaptiveCpp deferred task launch} 

In addition to in-order queues with explicit synchronizations, the SYCL standard allows the use of out-of-order queues and data dependencies for scheduling. As the SYCL standard allows mixing both approaches in the same application, AdaptiveCpp, by default, does not directly forward the submitted tasks to the underlying runtime, but instead uses an internal task graph. 

This results in multiple layers of asynchrony: when the application calls the SYCL API to submit a task to the GPU, the AdaptiveCpp runtime places this task into the internal buffer and later calls the HIP API from a separate thread; and HIP API itself is asynchronous for most operations (and relies on the HSA backend), submitting them directly to the user mode queue on the GPU but not waiting for their completion.

By default, the AdaptiveCpp task graph starts submitting the operations to the GPU only when a synchronization API is called or when the number of cached tasks reaches a set threshold (by default, 100). This introduces a delay between the application submitting the operation (such as kernel launch) to the SYCL runtime and the operation being submitted to the HIP runtime (and, eventually, to the GPU). When GROMACS is running in fully-GPU-resident mode, the resulting kernel launch delay can exceed 500~\textmu{}s, since the synchronization with host is performed very rarely. In a heterogeneous run with CPU force contributions (see \figref{fig:dd-md-schedule}; not explored in depth in this work), the CPU-GPU synchronization is performed every step, and thus the introduced delay is less significant. In multi-GPU runs, since MPI has no interoperability with GPU queues, initiating MPI transfers from the host requires CPU-GPU synchronization at least twice per MD step (for coordinate and forces halo exchange), so the kernel launch delay is also not significant in absolute terms. However, the delayed kernels can be on the critical path, and thus even a small delay can harm the application performance.

Since AdaptiveCpp (hipSYCL) version~0.9.4, task caching can be controlled by the \texttt{HIPSYCL\_RT\_MAX\_CACHED\_NODES} environment variable (later referred to as \textit{MCN}), which sets the maximum number of task graph nodes (GPU operations) that can be cached before the task graph is flushed. Setting MCN to zero should, theoretically, be optimal, since GROMACS does not rely on any SYCL features benefiting from deferred execution. However, the effects are more complicated. Frequent task graph flushing increases the load on the AdaptiveCpp runtime due to the ``bookkeeping'' required on each DAG flush. This load manifests in higher CPU usage by both AdaptiveCpp worker threads and extra HIP API calls to check the execution status of the tasks.

Furthermore, such asynchronous submission and caching can hide some of the latencies of HIP API calls. For example, in our observations, \texttt{hipStreamWaitEvent} call can take up to 50~\textmu{}s\cite{GmxSycl:2023}. In the presence of CPU work, making this API calls from a separate thread and thus moving it off the critical path  can theoretically improve performance compared to the native runtime. Hiding the GPU API latencies from the main application thread can also prove advantageous when the task submission is not on the critical path, e.g., if GPU-initiated communications or stream-aware MPI \cite{StreamAwareMpi:2022,MpiStream:2022} are used.

In our benchmarks, we compare three different values of MCN: 0 (each operation is immediately submitted to the GPU from the separate worker thread), 5 (chosen to be slightly less than the typical number of API calls per MD step\footnote{The exact number of operations submitted to the GPU per step varies significantly based on the offloaded computation and the decomposition used.}), and 100 (the default value).

AdaptiveCpp~23.10.0 in instant submission mode and GROMACS~HIP do not use deferred execution, calling the HIP API directly from the main application thread, thus they are not affected in any way by AdaptiveCpp task graph caching.

\subsubsection{CPU usage by runtime}

The CPU/GPU performance ratio on Cray EX235a is heavily skewed towards GPU, with a single compute core complex (CCX), containing 8 physical cores, per GCD. This is further exacerbated by some sites, like LUMI and Frontier (but not Dardel), reserving one CPU core on each CCX to enable \emph{low-noise mode}, leaving only 7 CPU cores available to applications. This makes it crucial to optimize resource usage.

The AMD runtime spawns the HSA worker thread, present in all HIP applications. In our observations, it consistently has 50-100\% CPU usage.

In addition to the HSA worker thread, AdaptiveCpp in the default submission mode introduces two additional worker threads to manage its task graph and perform API calls to the GPU runtime. In AdaptiveCpp (hipSYCL) 0.9.4, the CPU usage of each thread can be over 80\% with MCN=0, requiring reserving two CPUs for the runtime. With the default MCN=100, the CPU usage of the worker threads is much lower, and both threads can share the same CPU. These overheads have been significantly reduced in AdaptiveCpp~23.10.0, with a single CPU now being sufficient even with low MCN values.

When the instant submission mode is enabled by defining the \texttt{HIPSYCL\_ALLOW\_INSTANT\_SUBMISSION=1} macro at compile time, AdaptiveCpp does not use its submission runtime, instead performing HIP API calls directly from the application thread. This mode, in our experience, typically achieves the best performance, but as noted before, asynchronous submission can be advantageous in the case of poor performance of the underlying GPU runtime or for applications with different scheduling patterns.

\subsubsection{GPU queue multiplexing}

When running on a system with multiple AMD GPUs, it is common to restrict the GPU visibility per-process using \texttt{ROCR\_VISIBLE\_DEVICES} or \texttt{HIP\_VISIBLE\_DEVICES} environment variables. This approach is more straightforward than application-specific custom solutions, especially with the complex CPU NUMA to GPU mapping on many machines, and it is often used in compute center documentation. This also has some unexpected benefits when using AdaptiveCpp. Until version 24.02.0, the AdaptiveCpp runtime created four idle \texttt{hipStream}s on each visible device. While these streams consume little resources, the presence of these HIP streams changes the mapping between HIP streams and hardware queues, and, in turn, affects whether the work in different HIP streams can overlap or not. In our experience, leaving all the devices visible and relying on the application-specific device selection logic can cause performance drops up to 10\%. This problem is avoided when only one device is visible to the application process (e.g., via \texttt{ROCR\_VISIBLE\_DEVICES}). Note that the ROCm \texttt{GPU\_MAX\_HW\_QUEUES} environment variable can affect how HIP streams are assigned to hardware queues \cite{AmdHwQueues:2018} and thus it affects the observed behavior, albeit in hard to predict ways. We have also observed a large drop in performance when using more than 5 GCDs per node, which is also resolved by restricting device visibility.

\subsubsection{Runtime overheads summary}

Reducing the MCN value has two opposite effects. On the one hand, frequent task graph flushes (with low MCN values) harm the performance of the SYCL runtime, especially in AdaptiveCpp (hipSYCL) 0.9.4. On the other hand, delaying the task graph flush (with high MCN values) can delay the kernel submission to the GPU. The balance of these two effects depends a lot on the runtime schedule. On a single GCD in fully GPU-resident mode, the CPU schedules work for multiple MD timesteps in advance. If the kernels are very small, we are primarily limited by the task submission, and thus runtime inefficiencies are noticeable, but the kernel launch delay is less critical since kernel time is anyway small. As the system size increases and the kernel wall-time grows, the kernel time plus the submission delay start to affect the application runtime; at this transient stage, reducing the runtime delay by lowering MCN can improve performance, even if it slows down the task submission. Finally, for very large kernels, the application performance is determined by kernel walltime, and runtime overheads and launch delay matter less.

The behavior gets more complex when we have multiple ranks with MPI communications (\figref{fig:dd-md-schedule} shows an example of such schedule), since now we have host-initiated communications that must wait for the completion of GPU work several times per time step, and thus the effects outlined above now apply to different groups of less than ten kernels each.

A further complication, not discussed in this work but observed in practice when having force computation on the CPU, is the competition between application threads and runtime threads for CPU resources.

All of these overheads (except the one coming from the HSA thread, which is part of the AMD ROCm runtime and is present in all HIP applications) are avoided by using the instant submission mode introduced in AdaptiveCpp~23.10.0. While it is not compatible with some SYCL features (notably, out-of-order queues and reductions), it is fully supported in GROMACS, and, with the current scheduling approach, tends to offer superior performance, as discussed in the next section.

\subsection{Single GPU}

To gauge the application performance on a single GCD, we used the set of Grappa PME inputs. These water-ethanol benchmark systems have uniform density and are easily scaled in size while still containing a mix of force workload representative of a biomolecular system (from a task wall-time point of view). 

\begin{figure}
    \centering
    \includegraphics[width=3.4in]{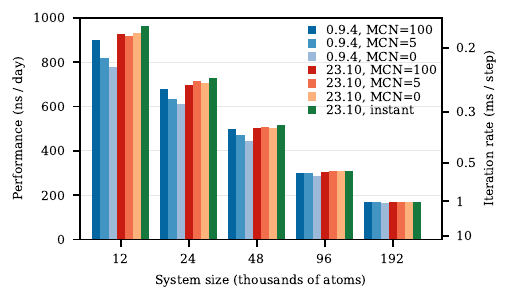}
    \caption{Application performance (ns / day) and corresponding iteration rate (ms / step) for a different Grappa PME systems on a single GCD. Performance of GROMACS~2024.0 compiled with different SYCL runtimes (AdaptiveCpp (hipSYCL)~0.9.4, AdaptiveCpp~23.10.0, and AdaptiveCpp~23.10.0 with instant submission mode enabled) is presented with various caching settings (MCN). The instant submission mode does not use caching. Median of five runs.}
    \label{fig:single-gpu-mcn}
\end{figure}

Figure~\ref{fig:single-gpu-mcn} shows performance results of a range of Grappa PME inputs. The results are presented for 5 application threads (in addition to AdaptiveCpp and HSA runtime threads) pinned to a single CCX (8 cores, 16 hardware threads), leaving three full cores to the runtime/driver to minimize any effects of the CPU contention. GROMACS is running in fully-GPU-resident mode, meaning the application schedules the GPU work for almost a hundred steps in advance, then waits for these to complete.

On the left side of the figure, the application is latency-bound and the kernels cannot fully saturate the GPU, making the application performance sensitive to any runtime overheads. For a system of 12 thousand atoms, the iteration rates are around 0.2~millisecond. With AdaptiveCpp (hipSYCL) 0.9.4, the best performance (902~ns/day) is achieved with MCN=100.
More frequent task graph flushing increases the associated overheads, reducing the performance by 14\% for MCN=0 despite potentially eliminating the kernel launch delay.
In AdaptiveCpp~23.10.0, the graph flushing overheads are less severe and the performance ranges from 921~ns/day (MCN=5, worst) to 932~ns/day (MCN=0, best).
The instant submission mode, which incurs minimal runtime overheads, improves the performance further, to 964~ns/day.
The effect of the runtime choice diminishes for larger systems. For the 24-thousand-atom system, the performance difference between the best and the worst MCN setting of AdaptiveCpp (hipSYCL) 0.9.4 is 10\%; and AdaptiveCpp~23.10.0 is 5--7\% faster.
With 192 thousand atoms, the differences between different runtime versions are even smaller: the best is AdaptiveCpp~23.10.0 in instant submission mode (171.5~ns/day), and the worst is 0.9.4 with MCN=0 (167.2~ns/day), less than 3\% difference.

\subsection{Multi-GPU scaling}

Scaling a system with PME to multiple nodes using distributed memory and message-passing is limited by the FFT scalability primarily due to limitations of the FFT libraries on AMD GPUs.
Exploring the performance of a multi-node distributed FFT is outside the scope of this work, thus we are using Grappa RF system, which uses reaction-field electrostatics and does not need PME.

When scaling to multiple GPUs, not only do we have less work per rank, but we also have a significantly different schedule. While GPU-aware MPI can use DMA between the GPU and NIC to avoid the need to stage the data in the main CPU memory, it still requires initiating the transfers from the host, introducing at least two synchronizations between CPU and GPU per MD step (see \figref{fig:dd-md-schedule}). Thus, GPU work can not be scheduled for multiple iteration in advance, the AdaptiveCpp task graph is flushed much more often, and the effects of delayed kernel launch are very different.

\begin{figure}
    \centering
    \includegraphics[width=3.4in]{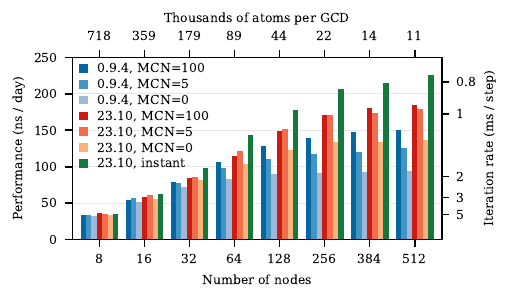}
    \caption{Application performance (ns / day) and corresponding iteration rate (ms / step) for a 46M Grappa RF box on multiple LUMI nodes. Performance of the GROMACS~2024.1 compiled with different SYCL runtimes (AdaptiveCpp (hipSYCL) 0.9.4, AdaptiveCpp 23.10.0, and AdaptiveCpp 23.10.0 with instant submission mode enabled) is presented with various caching settings (MCN). The instant submission mode does not use caching.}
    \label{fig:lumi-mcn}
\end{figure}

Figure~\ref{fig:lumi-mcn} presents the strong-scaling performance of a large Grappa RF system up to 512 nodes.

At the scaling limit, there are just 11 thousand atoms per GCD, and the iteration time is below 1~millisecond, with multiple MPI communications per step. For a 12 thousand atom system on a single GPU, the benefits of the instant submission were around 3\% (compared to the optimal cached performance), but in the multi-GPU run, the instant submission mode is 22\% faster than the fastest cached mode (226~ns/day vs. 185~ns/day). We also see much stronger effect of runtime overhead associated with frequent graph flushing: flushing every step reduces the performance, compared to the default behavior of MCN=100, by 38\% for AdaptiveCpp (hipSYCL) 0.9.4 and by 26\% for AdaptiveCpp~23.10.0.
As for the single-GCD case, the runtime effect diminishes when we have more work per GCD, but even at 192~thousand atoms per GCD the runtime effects are significant: the instant submission mode is 15\% faster than the fastest cached mode, while the same difference is under 1\% for a 192-thousand-atom system on a single GPU.
We can also observe that when the GPU work gets large enough, the MCN=5 case starts outperforming the MCN=0, meaning that the effects of deferred kernel launch are larger than the effects of graph flushing overheads. For AdaptiveCpp~23.10.0 that happens at the iteration rate around 1~millisecond (256 nodes, 22 thousand atoms per GCD), and for AdaptiveCpp (hipSYCL)~0.9.4 the threshold is around 3~millisecond (16 nodes, 359 thousand atoms per GCD).

\section{Performance results}

\subsection{Single-rank performance}

To evaluate the GROMACS performance on a single GCD, we used two common small- to medium-sized benchmark systems: RNAse and ADH.

\begin{figure}
    \centering
    \includegraphics[width=3.4in]{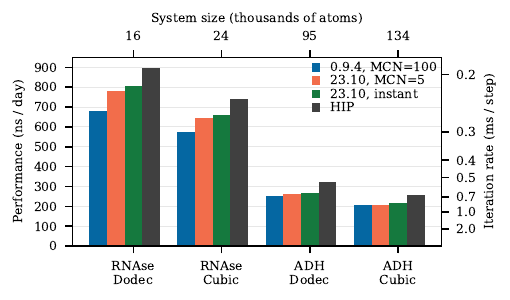}
    \caption{Application performance (ns / day) and corresponding iteration rate (ms / step) for benchmark systems, RNAse and ADH, running on a single GCD on Dardel. Performance of GROMACS~2024.0 compiled with different SYCL runtimes (AdaptiveCpp (hipSYCL)~0.9.4, AdaptiveCpp~23.10.0, and AdaptiveCpp~23.10.0 with instant submission mode enabled), compared to the performance of the HIP fork.}
    \label{fig:single-gpu-all}
\end{figure}

Figure~\ref{fig:single-gpu-all} contains the performance results (median of three runs), using 7 application threads (in addition to AdaptiveCpp and HSA runtime threads) pinned to a single CCX (8 cores, 16 hardware threads). Since GROMACS is running in fully-GPU-resident mode, there is little overlap between CPU work and GPU work, and using 7 worker threads in GROMACS works reasonably well across all configurations presented (with one exception noted below).
However, if there was CPU force computation, both application threads and GPU runtime threads would be active at the same time, and then the optimal number of application threads would depend much more on the GPU runtime version. For the systems presented here, the optimal MCN values were 100 for AdaptiveCpp (hipSYCL) 0.9.4 and 5 for AdaptiveCpp~23.10.0.

ADH systems running on a single GCD have an iteration time around 0.5--0.7~ms. They are not very susceptible to the runtime performance, but the difference is still observable. For ADH Dodec with AdaptiveCpp (hipSYCL) 0.9.4, the best performance is with MCN=100. Runtime optimizations in AdaptiveCpp~23.10.0 allow achieving 4\% higher application performance. The GROMACS~HIP performance is 22\% higher. For a slightly larger ADH Cubic system, the performance improvements when upgrading from AdaptiveCpp (hipSYCL) 0.9.4 to 23.10.0 is 3\% (from 209.3~ns/day to 215.3~ns/day when using MCN=0; MCN=5 is not optimal here for AdaptiveCpp~23.10.0). The instant submission mode yields another 1\% improvement; GROMACS~HIP version has 19\% higher performance.

A much smaller RNAse system has higher iteration rates, 0.2--0.3~ms, and thus the dependence on the runtime version is much more noticeable. AdaptiveCpp (hipSYCL) 0.9.4 has highest runtime overheads and lowest performance. Furthermore, the CPU competition between GROMACS and SYCL runtime is non-negligible, and reducing the number of application threads from 7 to 5 can further improve performance by 3\% to 4\% (data not shown). With AdaptiveCpp~23.10.0, the performance increases by 12\% for the larger Cubic system and  by 15\% for the smaller Dodec system. The instant submission mode, which does not use any additional runtime threads, improved the performance further by 3\% in both cases. The HIP fork is 11--12\% faster.

This difference of 10--20\% observed between the SYCL version and the HIP version is in line with the difference in the NBNXM kernel performance (see \figref{fig:kernels-relative}), which is known to be the bottleneck in single-GPU runs, and is expected to be bridged as more optimizations are ported and the AMD GPU compiler issues are worked out so less device-specific workarounds are necessary.

\subsection{Multi-GPU strong-scaling performance}

\begin{figure}
    \centering
    \includegraphics[width=3.4in]{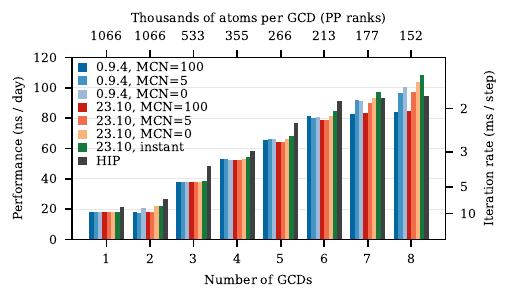}
    \caption{Application performance (ns / day) and corresponding iteration rate (ms / step) for the STMV benchmark running on different number of GCDs within a single node. Performance of GROMACS~2024.0 compiled with AdaptiveCpp (hipSYCL)~0.9.4 and AdaptiveCpp~23.10.0 is presented with various caching settings (MCN), as well as performance of GROMACS~2024.0 with AdaptiveCpp~23.10.0 with instant submission mode and the GROMACS HIP fork.}
    \label{fig:single-node-stmv}
\end{figure}

The STMV benchmark system, containing one million atoms, is a good use-case for single-node scaling, since its size allows it to both fit onto a single GCD and comfortably scale up to 8 GCDs (with 152 thousand atoms per each PP rank, and one dedicated PME rank). While not covering extreme values, it still shows the wide range of runtime behaviors, as presented in \figref{fig:single-node-stmv}, which compares the performance of GROMACS~2024.0 in different configurations and the HIP fork.

When running on a single GCD, the application performance is limited by the performance of force compute kernels, primarily NBNXM (of ``tabulated electrostatics, force-switch Lennard Jones'' flavor). This results in all SYCL versions having approximately the same performance of 17.8~ns/day, since the differences in SYCL runtime behavior do not manifest at the iteration rate of many milliseconds. The HIP version is 21\% faster, at 21.6~ns/day. The total kernel time (serialized) of SYCL version is 25\% higher than with HIP, which explains the observed differences in the total application performance.

When running on more than one GCD, we dedicate one rank to the PME (long-range) work and the rest to the PP (short-range) work. Therefore, when going from 1 to 2 ranks, the NBNXM kernels, which are rate-limiting, are still fully running on single GCD, and the performance gain compared to one GCD is rather small\footnote{In some cases, PME rank can be placed on the same GPU as one of the PP ranks to achieve better performance with the same number of GPUs, but this case is not explored here.}. However, the differences in the SYCL runtime behavior begin to manifest. With GPU kernel having significant walltime, the launch delay introduced by task caching, coupled with frequent device-host synchronizations to initiate MPI transfers, drastically reduce the performance. The runs without task caching have performance of 21.8--21.9~ns/day (AdaptiveCpp 23.10.0 with MCN=0 and the instant submission mode, respectively), around 23\% higher than with any other caching setting (17.6--17.9~ns/day with AdaptiveCpp 23.10.0). The HIP fork has performance of 26.8~ns/day, 22\% faster than the fastest SYCL version.

Going from 2 to 3 GCDs, we are splitting the NBNXM work to two GCDs, resulting in around 75\% speedup despite only 50\% more compute resources. When scaling to 7 or more GCDs, the single PME rank becomes the performance bottleneck. At this stage, NBNXM kernels become less relevant, with application performance determined by PME computation and communication. As GROMACS~2023 and newer contain numerous optimizations to its MPI communication patterns\footnote{One optimization in particular, \url{https://gitlab.com/gromacs/gromacs/-/merge_requests/3542}, is the main reason for the observed difference; disabling it reduced GROMACS 2024.0 performance on 8 GCDs by around 15\%.}, it continues scaling up to 8 GCDs, while the performance of the HIP fork plateaus after 6 GCDs. Scaling the performance past a single node would require the use of PME decomposition, which, in GROMACS~2024, has limited scalability on AMD GPUs\cite{Kronberg:poster:2024}.

\subsection{Multi-node strong scaling performance}

\begin{figure}
    \centering
    \includegraphics[width=3.4in]{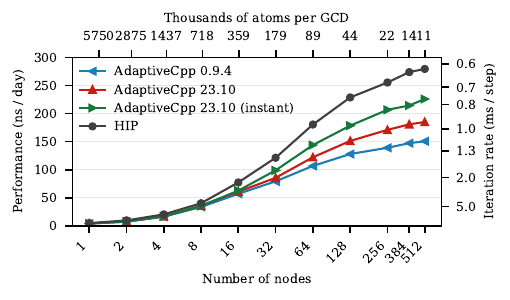}
    \caption{Application performance (ns / day) and corresponding iteration rate (ms / step) for a 46M Grappa RF system on multiple LUMI-G nodes. Performance of the GROMACS 2024.1 compiled with different SYCL runtimes (AdaptiveCpp (hipSYCL) 0.9.4, AdaptiveCpp 23.10.0, and AdaptiveCpp 23.10.0 with instant submission mode enabled) is compared with the performance of the HIP fork of GROMACS. For AdaptiveCpp, where applicable, the most optimal caching setting was chosen (see \figref{fig:lumi-mcn} for detailed analysis of its impact).}
    \label{fig:lumi-multinode-scaling}
\end{figure}

To assess the performance of GROMACS 2024 at scale, we carried out benchmarks on up to 512 nodes of LUMI-G. We evaluate the strong scaling of the particle domain decomposition halo exchange (without the PME long-range component) using a 46M atom Grappa RF system. The simulation use GPU-resident parallelization with all tasks offloaded. We compare the three major AdaptiveCpp runtime variants also showing the AMD HIP prototype fork performance in \figref{fig:lumi-multinode-scaling}.

As discussed earlier, the AdaptiveCpp instant submission compared to caching shows a clear benefit in MPI strong scaling starting from 16 nodes (3\%) increasing to 22\% on 512 nodes, as well as significantly better parallel efficiency.
While scaling with AdaptiveCpp caching effectively stops past 256 nodes, with instant submission strong scaling continues up to 512 nodes, corresponding to just 11k atoms per GCD and under 0.8~ms/iteration. We observe over 37\% parallel efficiency up to 128 nodes (44k atoms per GCD) from where scaling deteriorates faster. This is expected since the NBNXM pair interaction kernel, which entirely dominates the compute work here (since there is no PME or FFT computation), starts to decline in efficiency below 100 thousand atoms per GCD as shown in \figref{fig:kernels-absolute}. Similar scaling behavior has been observed on other platforms.
The parallel efficiency of GROMACS~2024.1 with AdaptiveCpp instant submission is on par with that of the HIP fork.
While absolute performance is 12--22\% lower, the majority of this difference is due to the performance gap in the NBNXM kernel(see \figref{fig:kernels-relative}), to a large extent caused by tail-effects due to different workload sorting. This affects the non-local short-range kernel throughout the entire scaling regime (given the small halo work), while both local- and non-local are affected at high parallelization. With the ongoing work of porting the improved sorting to SYCL, we expect to reduce the performance gap throughout the scaling range, and in particular above 32--64 nodes. GPU queue multiplexing issues likely impact scaling, and might do so differently for the different runtimes and configurations benchmarked.

A wide range of biomolecular MD studies require PME long-range forces, but multi-node strong scaling of such simulations is currently limited on the AMD platforms targeted here due to the the lack of scalable distributed 3D FFT library (suitable for our problem sizes). This makes the most common sized biomolecular simulations (typically less than 1 million atoms in size) impossible (or impractical) to scale beyond a single node.

\section{Future work}

Having tackled many of the GPU runtime challenges, two major steps remain for better strong-scaling performance of GROMACS: the kernel performance and the FFT scaling.

The kernel performance (primarily NBNXM) is the main factor for the performance difference between the HIP and the SYCL backend when running simulations without PME electrostatics. Implementing post-pruning pair list sorting to reduce tail effects of the NBNXM kernels is currently in progress and is planned for  inclusion in GROMACS~2025. The fine-tuning of kernel parameters is another important step, however it is stymied by the rapid development of the HIP compiler: e.g., we observed up to 20\% difference for the same kernel code between different compiler versions in ROCm 5 branch (data not shown).

A system with PME electrostatics can hardly be scaled past 8 GCDs with only on PME rank. The efficiency of the heFFTe library for small grids is much lower than that of the CUDA backend with cuFFTMp, drastically limiting the PME strong scaling. Vendor-tuned distributed FFT libraries optimized for our problem sizes on AMD GPUs would be preferred. However, some improvements here could come both from tuning of FFT use in GROMACS, as well as from improvements in heFFTe itself.

In addition to this directions, there are still a lot of possible improvements to the runtime performance. We have e.g. observed that the \texttt{GPU\_MAX\_HW\_QUEUES} setting of the HIP runtime can have a significant effect on the application performance when running multiple ranks, but the effect is non-linear, and lower values can lead to improved performance.

Recently, AMD began work on fully integrating the HIP backend into GROMACS, presenting an opportunity to evaluate the merits and limitations of various approaches within a highly optimized codebase, with important lessons for long-term performance portability work. The collaboration with AMD could benefit all backends by leveraging vendor expertise to address API-independent challenges, such as enhancing the scalability of distributed FFT libraries or making the ROCm runtime behavior more predictable.

The drawback of current GPU-aware MPIs that use CPU-initiated communication and lack GPU queue-awareness is known to impact scalability. GPU-initiated communication has shown promising results in improving FFT scalability\cite{cuFFTMp-blog:2022} and early results have been promising for the GROMACS halo exchange as well\cite{Gray2023GTCtalk}. This will require a stable and performance vendor library: rocSHMEM is still experimental and the HPE Cray software stack does not provide a GPU-resident SHMEM implementation.
Stream-aware MPI\cite{StreamAwareMpi:2022} also promises to remove the MPI-related waits from the critical path, but its implementation is not yet available on any of the systems used in this work.   

\section{Conclusions}

In this work, we demonstrated that SYCL can be an effective, performance-portable programming model for targeting AMD GPU-based HPC platforms. By building on the same native toolchains as vendor-specific languages like HIP, SYCL can achieve kernel performance comparable with non-portable approaches, as long as limited device-specific optimizations is accepted.

The additional abstraction layer introduced by SYCL can lead to runtime overheads that harm performance of latency-sensitive scenarios prevalent in molecular dynamics and other strong-scaling applications, often struggling even when using AMD ROCm directly\cite{EarlyMdAmd:2022}. However, this can be mitigated by the use of extensions or custom, platform-specific settings. We present a number of possible solutions for the AdaptiveCpp SYCL runtime, developed in collaboration between the GROMACS and  AdaptiveCpp teams.

Overall, the SYCL-based approach has proven viable for GROMACS on AMD GPUs. Although it required significant initial effort and close collaboration with the AdaptiveCpp developers, the benefits of avoiding code duplication and improved portability outweighed the challenges. Using SYCL can even lead to better-than-native performance in some cases by enabling the easier integration of optimizations initially aimed at other backends. And since the resulting runtime improvements are generic, they will benefit other projects using AdaptiveCpp.

A portable abstraction layer reduces long-term maintenance burden and eases extension to new hardware, even for a notoriously complex MD problem. There are features in native APIs not yet present in portable ones, but the interoperability in SYCL allows to use them selectively, minimizing code duplication and allowing a finer balance between performance and maintainability. As the diversity of hardware landscape is growing, the performance-portable approaches could be much more viable than maintaining multiple vendor-specific backends.

While the analysis in this paper is limited to GROMACS, we expect the techniques and lessons learned to be applicable to a broad range of other applications targeting heterogeneous HPC platforms. With further optimizations and improvements in distributed FFT performance, we anticipate being able to achieve even better performance at scale.

\section*{Acknowledgment}

We thank 
Aksel Alpay (Heidelberg University) for the fruitful discussions on improving the runtime performance and the implementation of the necessary optimizations in AdaptiveCpp;
Maciej Szpindler (LUMI User Support Team) for help with collecting strong scaling data and in-depth testing on LUMI-G;
Ragnar Sundblad (PDC Center for High Performance Computing) for help with node allocation on Dardel;
Julio Maia (AMD) for the suggestions on kernel optimizations;
Bálint Soproni (Stream HPC and Heidelberg University) for contributing VkFFT support to GROMACS and the suggestion on kernel optimizations;
and all contributions from the greater GROMACS community.
PDC Center for High Performance Computing and LUMI, and AMD are kindly acknowledged for hardware access and support.

\section*{Supplementary information}

The benchmark topologies and input parameters can be downloaded from \url{https://doi.org/10.5281/zenodo.11087335}.

\bibliographystyle{IEEEtran} 
\bibliography{IEEEabrv,bibliography.bib}

\end{document}